# Increase of the electromechanical coupling of piezoelectric vibration harvesters through lateral bars

**David Gibus[1,*], Grégoire Forges[1,2], Hélène Debéda[2], Pierre Gasnier[3] and Adrien Badel[1]**

[1] SYMME, Université Savoie Mont-Blanc, F-74000, Annecy, France
[2] Université de Bordeaux, CNRS, Bordeaux INP, IMS, UMR 5218, 33400 Talence, France
[3] Université Grenoble Alpes, CEA, LETI, MINATEC, F-38000 Grenoble, France

*Author to whom any correspondence should be addressed.

E-mail: david.gibus@univ-smb.fr

## Abstract

To enhance the performance of vibration energy harvesters, it is essential to maximise the electromechanical coupling coefficient $k^2$ of piezoelectric devices. This enables sufficient harvested power and tuning capability of the resonant frequency through electrical methods. While much literature treats the optimisation of piezoelectric cantilevers, the optimisation range is usually limited by the transverse coupling coefficient $k_{31}^2$ of the material. This work introduces an innovative solution to extend the optimisation range and increase the coupling coefficient of piezoelectric cantilevers. This is achieved by minimising lateral strain in the beam using lateral bars to maximise the equivalent material coupling coefficient.
The theoretical basis of this innovation is demonstrated through the exploitation of the constitutive equations of piezoelectric materials. The interest of the addition of lateral bars to increase the coupling coefficient is demonstrated and studied with simulations based on the finite elements method.
Finally, a proof of concept is realised using a aluminum cantilever prototype integrating a lead-free $KNaNbO_3$ (KNN) piezoelectric material. It is tested under vibration at 0.1 m/s² with variable resistive loads. The results show that the coupling coefficient increases by 30% and the relative frequency bandwidth by 32% with resistive tuning of the resonant frequency, by fixing eight steel bars to the cantilever. The designed prototype is a highly performant lead-free vibration energy harvester. It produces a maximum power of 49.9 µW at resonance, and its normalised power density is equal to 16.5 mW/G²/cm³. Its relative frequency bandwidth is equal to 3.1%.

---

## 1. Introduction

Among piezoelectric devices, the piezoelectric cantilever is a widely used and highly effective structure for many applications [1,2]. They are used in applications ranging from vibration energy harvesting [3,4] to actuators [5,6] and vibration mitigation [7]. Their interest lies in their simple design, which allows for the production of low-resonant-frequency structures with good electromechanical coupling coefficients, exploiting the $k_{31}$ coupling coefficient of materials. Resonant frequencies can easily be adjusted, and they are easy to produce. Various works have been proposed for optimization purposes. Among the design considerations, maximizing the electromechanical coupling coefficient $k^2$ is one of the most critical parameters. Indeed, when designing piezoelectric vibration energy harvesters, maximizing $k^2$ is necessary to maximize the power harvested for weakly to moderately coupled piezoelectric harvesters [8]. Moreover, maximizing $k^2$ improves the tuning range of the resonant frequencies via electrical methods for strongly coupled harvesters [9]. Several electrical methods and implementations have indeed already been proposed for tuning the resonant frequency of piezoelectric vibration energy harvesters [10]. When working to maximize the coupling coefficient $k^2$ of a cantilever there are two optimisations to consider: i. Using piezoelectric materials with strong electromechanical coupling coefficients, ii. Optimising the mechanical design of the device to concentrate elastic energy in the piezoelectric material in the most uniform way [11]. Lead-based piezoelectric materials exhibit strong coupling coefficient (e.g., PZT, PZN-PT, or PMN-PT), enabling high global coupling coefficients ($k_{31}^2 > 20$ % and $k^2 > 10\%$) to enhance the frequency tuning capability [9,10]. However, in order to mitigate the environmental risks posed by lead, it is essential to optimize lead-free vibration energy harvesters. Lead-free piezoelectric materials (e.g., $KNaNbO_3$ (KNN) or $LiNbO_3$) have generally lower coupling coefficient than lead-based counterparts, and the electromechanical coupling coefficient $k^2$ of the designed harvesters rarely exceeds 2% [12–14]. It is therefore essential to implement design optimizations to increase the electromechanical coupling of structures made from lead-free materials or to further improve the performance of those made from lead-based materials. This work introduces a solution to increase the electromechanical coupling capability of the piezoelectric material in cantilevers through a mechanical innovation.

Design solutions have already been discussed in the literature, primarily focusing on the optimization of geometrical parameters. In 2021, Zhao *et al.* optimized the thicknesses and materials of the substrate and piezoelectric elements of their cantilever for frequency tuning in vibration energy harvesting applications [15], in the same way that Wang et al. did in 1999 for actuation applications [6]. They designed a wide prototype with a coupling coefficient $k^2$ of 14.0 % using PZT-5H material ($k_{31}^2 = 15$ %). Using a lead-free material, Clementi *et al.* achieved a coupling coefficient $k^2$ of 2.7% by optimizing the cut orientation of a $LiNbO_3$ single crystal ($k_{23}^2 = 24\%$) and the cantilever dimensions through FEM simulations [12]. Ducarne *et al.* [16] and Lossouarn *et al.* [16] conducted research on the placement and thickness optimization to improve the electromechanical coupling of several resonant modes. Design innovations have also been proposed in the literature to homogenize the strain in the beam. To this end, width-tapered beams [17,18] and thickness-tapered beams [19] have been shown to be effective in homogenizing the strain and increasing the electromechanical coupling coefficient of harvesters. Furthermore, the use of a long proof mass at the end of the cantilever tip has been shown to be a simpler and more efficient solution for strain homogenization and improving the coupling coefficient [20,21]. In 2022, a PZN-PT-based cantilever with a long proof mass was designed reaching coefficient $k^2$ of 49.8 % [9].

In addition to geometric optimization, work has been undertaken to exploit other material couplings than the conventional coupling $k_{31}$ in the beams. Devices were designed to exploit the coupling constant $k_{15}$, which is typically greater than $k_{31}$ , by inducing shear stress in the piezoelectric material [22]. However, their implementation is complex, and no advantage has been demonstrated in terms of the global coupling coefficient $k^2$. As complex structures are designed to generate shear stress in the piezoelectric material, deformations appear to occur more in the non-piezoelectric material than in the piezoelectric material. Furthermore, exploiting the coupling $k_{15}$ requires thick patches. Another possibility is to exploit the coupling $k_{33}$ withbeams with interdigitated electrodes proposed in the literature [23,24]. The electrode configuration allows the electric field to be oriented in the same direction as the strain in the beam. However, since the electric field is not entirely longitudinal, the coupling coefficients utilized are often lower than $k_{33}$ [25,26]. Furthermore, the capacitance is often low, which can pose energy management challenges for the circuit [24]. Another idea to make use of the coupling $k_{33}$ was suggested by Ahmed-Seddik *et al.* [27]. They designed a tapered cantilever with PZN-PT stacks ($k^2 = 49.3$ %), but the integration of multiple piezoelectric elements in mode 33 along the beam appears to significantly complicate its manufacturing and tends to reduce its mechanical robustness. More recently, a really simple way to maximize the coupling coefficient piezoelectric cantilevers was proposed by maximizing the width-to-length ratio of beams with piezoelectric ceramic [21]. When piezoelectric ceramics are used, the coupling coefficient is indeed maximized when the lateral strain is minimized by the exploitation of the coupling $k_{31}^w$ [28]. In [21], it has been shown that multiplying the width of PZT-5A-based cantilever by 5 increased the coupling $k^2$ by 45%, from 11.3% to 16.4% ($k_{31}^2 = 15$ % and ${k_{31}^w}^2 = 34$ %). However, the design of wide beams increases the complexity of the design when the overall volume is constrained. Respecting a width to length ratio superior to 5 is complex when the length of the beam has to be long enough to have low resonant frequency. Minimizing lateral strain could, however, be realized through mechanical innovation. The recent literature on auxetic devices may be perceived to minimize the lateral strain to exploit $k_{31}^w$, or to reverse the strain with a negative Poisson ratio to exploit the coupling $k_p$ of the radial mode [29]. However, the coupling of the cantilevers is rarely mentioned in the studies about auxetic structures, which does not permit a comparison of results [30]. Although the use of auxetic substrates [31] and auxetic boosters [32] could be efficient to engineer the lateral strain, they are complex to design and produce, and they induce a strong inhomogeneity in strains. Recently, the authors presented an innovative way to reduce the lateral strain of cantilevers and increase their coupling coefficient : the use of lateral bars [33]. Simulations have shown that attaching rigid steel bars in the width direction of the beam above the piezoelectric PMN-PT single crystals minimizes lateral deformation and maximizes coupling coefficient, increasing $k^2$ by 89 %. These bars appear to be a competitive solution for improving coupling, given their effectiveness and ease of implementation. It allows a significant increase in the coupling associated with simplicity of manufacture, all in a reduced space, since it is not necessary to have a structure of great width. However, its interest with conventional ceramic piezoelectric material (PZT or KNN) has never been discussed and an experimental validation is missing in literature. The results from [33] where only numerical results with a PMN-PT single crystal, which is really compliant and had a strongly coupling coefficient ${k_{31}^{w}}^2$ compared to $k_{31}^2$. The equivalent longitudinal stiffness of PMN-PT single crystal is around 20 GPa, while piezoelectric ceramics stiffness is usually greater than 60 GPa, and the coupling coefficient ${k_{31}^{w}}^2$ is 3.5 times greater than $k_{31}^2$ for the PMN-PT single crystal, while this ratio is generally around 2 for piezoelectric ceramics. Furthermore, since the results were numerical, it remained unclear whether it was possible to achieve a genuine reduction in lateral strain in the experiments and an increase in the coupling.

This article discusses the solution involving the placement of multiple rigid bars on piezoelectric ceramic materials to improve the coupling coefficient of narrow cantilevers, and presents the experimental validation. The merits of this solution are demonstrated through analytical modelling, and finite element simulations are performed to study its implementation. Parametric analyses identify the main design rules. A KNN-based prototype with an aluminium substrate is designed and tested to demonstrate the effectiveness of the bars in maximizing the coupling coefficient. The following section presents the design innovation and the piezoelectric modelling. Section 3 demonstrates the feasibility of maximizing the coupling coefficient for a cantilever configuration using COMSOL simulations. The final section presents experimental validation using a piezoelectric vibration energy harvester based on the KNN ceramic and equipped with frequency tuning capability, tested under vibration conditions.

## 2. Device presentation and modelling

### *2.1 Device presentation*

The studied cantilever is a piezoelectric bimorph with a proof mass (Figure 1). The two piezoelectric layers and the substrate have the same length ($L_b$) and the same width ($B$). As it has been shown as optimal configuration [21], the proof mass is long to homogenize the strain distribution in the beam. The piezoelectric layers are entirely covered with electrodes. The proof mass length and height are noted $L_m$ and $H_m$.

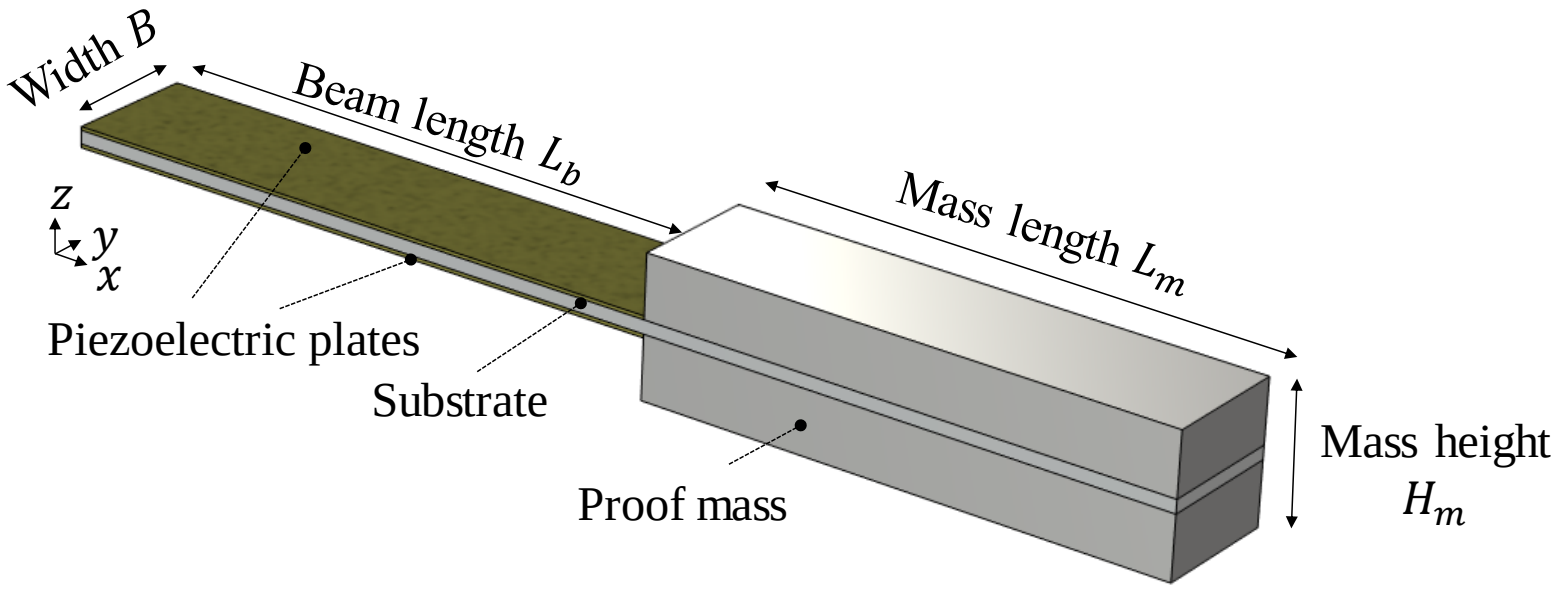

Figure 1: Piezoelectric cantilever with proof mass

The dimensions piezoelectric cantilevers are usually calculated to achieve the target frequencies of the applications. In general, the targeted frequencies are low (often below 100 Hz) [34]. Since the thicknesses are constrained by manufacturing processes, the length of the beams is often adjusted to match the desired frequency. As an illustration, equation (1) give the expression of the natural frequency $f_n$ of a simple cantilever with proof mass modeled with a 1-DOF modeling, considering the bending static stiffness, negligible beam mass to the mass of the proof mass, and that the proof mass width is equal to the beam width [34]. $E$ is the Young's modulus of the beam material, $\rho$ is the density of the proof mass. As indicated by equation (1), $f_n$ is inversely proportional to $\sqrt{L_b^3}$.

$$f_n = \sqrt{\frac{Eh^3}{12L_b^3} \times \frac{1}{\rho L_m H_m}} \quad (1) \qquad S_{beam} = L_b \times B \quad (2)$$

Furthermore, as demonstrated further in section 2.2, it has recently been shown that reducing lateral strain in piezoelectric harvesters allows for maximizing considerably their global coupling coefficient [21]. This was achieved by maximizing the width-to-length ratio $B/L_b$ to satisfy the assumption of lateral plane strain. Nevertheless, the design of wide beams is challenging if devices with small volumes and low resonant frequencies are required. Indeed, the desired $B/L_b$ ratio should be higher than 5 to fully satisfy the plane strain assumption. As the surface of the beam alone given in equation (2) depends on the beam width, increasing $L_b$ to decrease the natural frequencies while keeping a high width-to-length ratio increases considerably the harvester volume.

The Figure 1 shows how the beam surface that would change if $L_b$ were increased to reduce the natural frequency of the beam to a target value using equations (1) and (2) . The initial case corresponds to a natural frequency, noted $f_0$, and a surface $S$. Two cases are considered: i. The width is independent of the beam length and only $L_b$ is changed in the computation of the beam surface, ii. The width is proportional to $L_b$ and $S_{beam}$ is proportional to $L_b^2$. In Figure 2, we observe that the surface of the beam becomes important since its width is proportional to the length. As an example, to divide the frequency by 10, the length should be multiplied by 4.6 and the surface by 21.5. This outcome therefore justifies the prevalence of long and narrow beams being designed as vibration energy harvesters, as evidenced in the review of Sadaf *et al.* [35] and in the work of Peralta *et al.* when the size of the beam is considered as an optimization criteria [36].

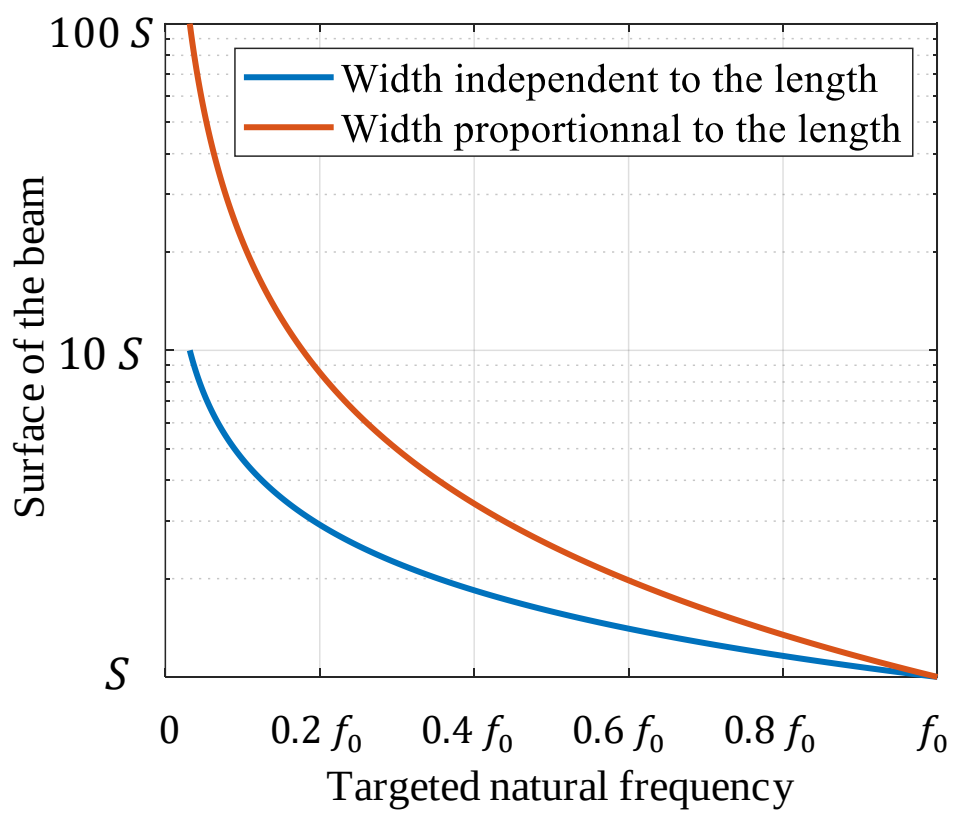

Figure 2: Influence of the resonant frequency on the beam surface

As a result, a harvester corresponding to the plane strain condition and having a low resonant frequency, which requires a large length $L_b$ and a very large width $B$, is challenging in a small volume. The objective is to restrict lateral deformations while enabling longitudinal deformations to achieve a high coupling coefficient with a minimized space requirement (small width). This is implemented by placing several rigid bars on the piezoelectric patches in the width direction. Figure 3 shows an enhanced version of a classic cantilever with the addition of bars on each side. The geometric parameters of the bars are the length, noted $l_b$, and height, noted $h_b$. The bars are considered covering all the width of the beam.

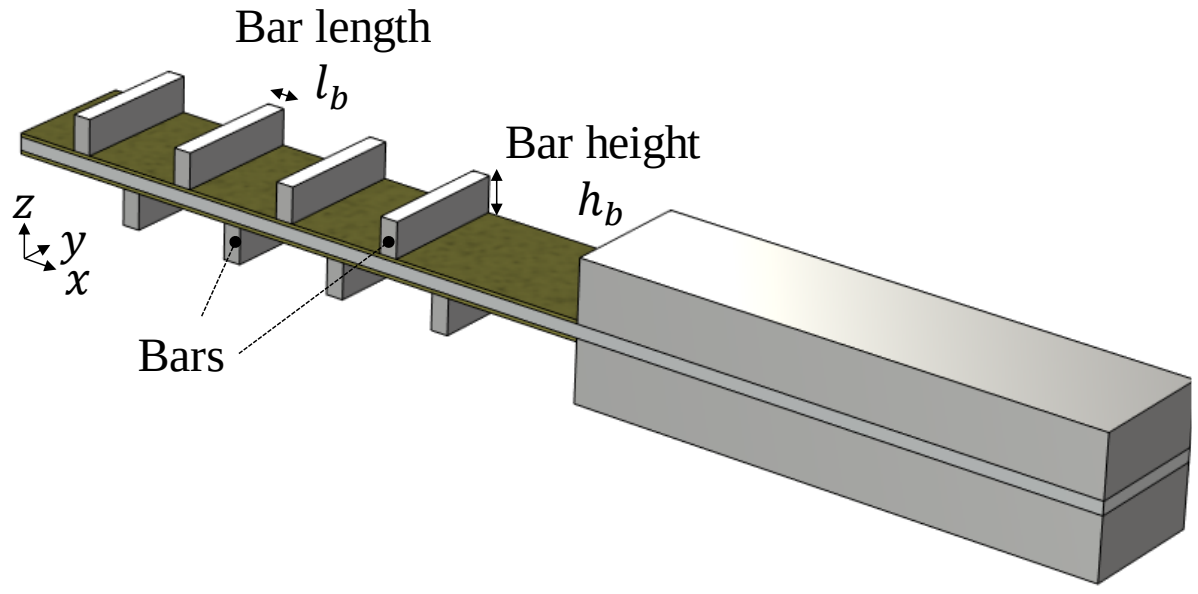

Figure 3: New design of the enhanced cantilever with metallic bars to increase the coupling coefficient $k^2$

As it will be discussed in the section 3, the bars must be wide enough to be rigid ($l_b$ must be large enough) and thin enough, not to prevent the beam from deforming in the longitudinal direction (the value of $l_b$ must not be too large). The bars are placed on the piezoelectric materials on both sides of the beam.

The improvement of the coupling depends on the "initial" cantilever: the coupling coefficients of the piezoelectric material, and its stiffness, the substrate nature and the geometry. The improvement also relies on the bars: their dimensions, the material stiffness as well as their number and arrangement (distance between each bar, staggered or face-to-face arrangement). As all the influences cannot be extensively analyzed, this work introduces a validation on an example configuration with a parametric study realized on the geometrical parameters of the bars.

The next section introduces the material modeling that justifies the importance of minimizing lateral strain. The constitutive equations for piezoelectric materials are used to demonstrate that coupling is improved when the strain in the direction of the beam's width is minimized. A finite element simulation validation is performed on a KNN-based cantilever with an aluminum substrate, and a parametric study of the bars is conducted in Section 3. Section 4 introduces the experimental validation on the KNN-based cantilever.

### *2.2 Rayleigh method*

This section introduces the modelling of piezoelectric cantilever using Hamilton principle and Rayleigh method. Considering the mode shape around the first resonant frequency, the equation of motion with one mechanical degree of freedom and the equation of charge are obtained. The expression of the electromechanical coupling coefficient $k_e^2$ around the resonant frequency is deduced from the equations and shows the dependency on the piezoelectric material solicitation.

The Hamilton principle [37] is applied to the structure (3). The virtual variations of the kinetic energy $T_k$, the internal electrical energy $W_e$, the potential energy $U$ and the external work $W$ are null between times $t_1$ and $t_2$. The proof mass is assumed to be rigid. The potential energy $U$ is equal to the sum of the energy stored in the piezoelectric material $U_p$, and that stored in the non-piezoelectric material $U_{np}$ (i.e. the substrate and the lateral bars), as shown by equation (4). In equations (7) to (10), the energy terms and the external work are deduced from the linear constitutive relations of piezoelectricity (5) and the Hooke law (6) ([38,39]). The external work $W$ correspond to the mechanical work due to the inertial forces $\boldsymbol{f_k}$ from the base acceleration and the relative beam displacement $\boldsymbol{u_k}$ and to the electrical work linked to the charge extraction as expressed in (11). $v$ is the voltage across the electrodes and $q$ is the extracted charge. The mechanical and dielectric losses are not considered in this stage of modelling.

$$\int_{t_1}^{t_2}[\delta(T_k - U + W_e) + \delta W)]dt = 0 \tag{3}$$

$$U = U_p + U_{np} \tag{4}$$

$$\begin{pmatrix}\boldsymbol{T}\\ \boldsymbol{D}\end{pmatrix} = \begin{bmatrix}\boldsymbol{c}^E & -\boldsymbol{e}^t\\ \boldsymbol{e} & \boldsymbol{\varepsilon}^S\end{bmatrix}\begin{pmatrix}\boldsymbol{S}\\ \boldsymbol{E}\end{pmatrix} \quad (5) \qquad\qquad \boldsymbol{T} = \boldsymbol{cS} \quad (6)$$

$$U_{sb} = \int_{\mathcal{V}_s} \boldsymbol{S}^t\boldsymbol{c_s}\boldsymbol{S} + \int_{\mathcal{V}_b} \boldsymbol{S}^t\boldsymbol{c_b}\boldsymbol{S} \tag{7}$$

$$U_p = \int_{\mathcal{V}_p} \boldsymbol{S}^t\boldsymbol{T}d\mathcal{V}_p \tag{8}$$

$$W_e = \int_{\mathcal{V}_p} \boldsymbol{E}^t\boldsymbol{D}d\mathcal{V}_p \tag{9}$$

$$T_k = \int_{\mathcal{V}_s} \boldsymbol{u}^t\rho_s\boldsymbol{u} + \int_{\mathcal{V}_b} \boldsymbol{u}^t\rho_b\boldsymbol{u} + \int_{\mathcal{V}_m} \boldsymbol{u}^t\rho_m\boldsymbol{u} + \int_{\mathcal{V}_p} \boldsymbol{u}^t\rho_p\,\boldsymbol{u} \tag{10}$$

$$\delta W = \sum_{k=1}^{n_f} \delta\boldsymbol{u_k}.\boldsymbol{f_k} + \delta v.q \tag{11}$$

Where $\boldsymbol{E}$, $\boldsymbol{S}$, $\boldsymbol{D}$ and $\boldsymbol{T}$ are the components of the electric field, strain, electric displacement and stress respectively. The superscript $(\,)^t$ denotes transpose. $\boldsymbol{u}$ is the relative displacement vector in the three direction $\boldsymbol{u}(\boldsymbol{x},t) = [u_x, u_y, u_z]$. $\boldsymbol{e}$ is the piezoelectric constant matrix, $\boldsymbol{\epsilon}^{\boldsymbol{S}}$ is the dielectric matrix at constant strain, $\boldsymbol{c}^{\boldsymbol{E}}$ is the elastic stiffness matrix of the piezoelectric material at null electric field. The subscripts $s$, $b$, $p$ and $m$ correspond to the substrate, the lateral bars, the piezoelectric material and the proof mass respectively. $\boldsymbol{c}$, $\mathcal{V}$, $\rho$ are the stiffness matrices, the volume and the densities respectively.

As electrodes cover the piezoelectric plates, the displacement field $\boldsymbol{D}$ and the electric field $\mathbf{E}$ are only considered in the transverse direction (i.e. $\boldsymbol{D} = D_3\vec{\boldsymbol{z}}$ and $\boldsymbol{E} = E_3\vec{\boldsymbol{z}}$ as $E_1 = E_2 = 0$ and $D_1 = D_2 = 0$). Furthermore, as the piezoelectric plates have the same poling direction and are connected in parallel, the electric field is in the same direction in the materials. The behaviour can be expressed as equation constitutive relations of piezoelectricity can be reduced to equation (12), where $\boldsymbol{e}$ is simplified as a vector as expressed given by $\boldsymbol{e} = [\, e_{31} \quad e_{31} \quad e_{33} \quad 0 \quad 0 \quad 0]$. If the piezoelectric plates are connected in series, rotation matrix should be considered [40].

$$\begin{cases} \boldsymbol{T} = \boldsymbol{c}^{\boldsymbol{E}}\boldsymbol{S} - \boldsymbol{e}^{\boldsymbol{t}}E_3 \\ D_3 = \boldsymbol{eS} + \epsilon_{33}^{S}E_3 \end{cases} \tag{12}$$

Using the Rayleigh approach to determine the behavior around the first resonant frequency, the $\boldsymbol{u}(\boldsymbol{x},t)$ can be expressed as the product of a mode shape $\boldsymbol{\phi}(\boldsymbol{x}) = [\phi_x(\boldsymbol{x}); \phi_y(\boldsymbol{x}); \phi_z(\boldsymbol{x})]$ multiplied by a generalized mechanical coordinate $r(t)$ as given in (13). The scalar electrical potential term, $\varphi$ can also be expressed via a potential distribution, $\Psi_v(\boldsymbol{x})$ and the generalized voltage coordinate, $v(t)$ as in equation (14).

$$\boldsymbol{u}(\boldsymbol{x},t) = \boldsymbol{\phi}(\boldsymbol{x})r(t) \tag{13}$$

$$\varphi(\boldsymbol{x},t) = \Psi_v(\boldsymbol{x})v(t) \tag{14}$$

In the same manner as in [40], the strain and the electric field is expressed using the differential operators $\boldsymbol{L_u}$ and $L_\varphi$ to the displacement and electrical potential in (15) and (16), respectively. The differentiated shape functions $\boldsymbol{N}$ and $N_v$ are defined in (18) and (17). While $\boldsymbol{N}$ is a vector ($\boldsymbol{N} = [N_1; N_2; N_3; N_4; N_5; N_6]$), $N_v$ is a scalar as only $E_3$ is considered. The differential operator $L_u$ is expressed in (19) from [41], and $L_\varphi = \partial/\partial z$.

$$\boldsymbol{S}(\boldsymbol{x},t) = \boldsymbol{L_u}(\boldsymbol{x})\,\boldsymbol{u}(\boldsymbol{x},t) = \boldsymbol{N}(\boldsymbol{x})\,r(t) \tag{15}$$

$$E_3 = L_\varphi\varphi(\boldsymbol{x},t) = N_v(\boldsymbol{x})v(t) \tag{16}$$

$$\boldsymbol{N}(\boldsymbol{x}) = \boldsymbol{L_u}\boldsymbol{\phi}(\boldsymbol{x}) \tag{17}$$

$$N_v(\boldsymbol{x}) = L_\varphi\Psi_v(\boldsymbol{x}) \tag{18}$$

$$\boldsymbol{L_u} = \begin{bmatrix} \frac{\partial}{\partial x} & 0 & 0 & \frac{\partial}{\partial y} & 0 & \frac{\partial}{\partial z} \\ 0 & \frac{\partial}{\partial y} & 0 & \frac{\partial}{\partial x} & \frac{\partial}{\partial z} & 0 \\ 0 & 0 & \frac{\partial}{\partial z} & 0 & \frac{\partial}{\partial y} & \frac{\partial}{\partial x} \end{bmatrix}^t \tag{19}$$

Applying the Rayleigh method [40], we obtain the equation of motion and the equation of charge that describe the behavior of the piezoelectric cantilever. $M$, $K_p$, $K_{np}$, $\Theta$, $C_p$ and $B_f$ are the equivalent mass, stiffness of the piezoelectric material, stiffness of the non-piezoelectric material, coupling term, capacitance and forcing term, respectively. Their expressions are given in equations (21) to (26). The forcing term is expressed considering the integral over the beam length and a mass per unit length $m(x_a)$ [42].

$$\begin{cases} M\ddot{r} + \left(K_p + K_{np}\right)r - \Theta v = -B_f\ddot{w}_B \\ \Theta r + C_p v + q = 0 \end{cases} \tag{20}$$

$$K_p = \int_{V_p} \boldsymbol{N}^t \boldsymbol{c}^{\boldsymbol{E}} \boldsymbol{N} \tag{21}$$

$$K_{np} = \int_{V_s} \boldsymbol{N}^t \boldsymbol{c_s} \boldsymbol{N} + \int_{V_b} \boldsymbol{N}^t \boldsymbol{c_b}\, \boldsymbol{N} \tag{22}$$

$$M = \int_{V_p} \boldsymbol{\phi}^t \rho_p \boldsymbol{\phi} + \int_{V_s} \boldsymbol{\phi}^t \rho_s \boldsymbol{\phi} + \int_{V_b} \boldsymbol{\phi}^t \rho_b \boldsymbol{\phi} + \int_{V_m} \boldsymbol{\phi}^t \rho_m \boldsymbol{\phi} \tag{23}$$

$$\Theta = \int_{V_{p1}} \boldsymbol{N}^t \boldsymbol{e}\, N_v \tag{24}$$

$$C_p = \int_{V_p} N_v^2 \epsilon_{33}^S \tag{25}$$

$$B_f = -\int_0^{L_b} m(x_a) \phi_3 dx_a \tag{26}$$

### *2.3 Electromechanical coupling coefficient*

From equation (20), the alternative coupling coefficient of the piezoelectric cantilever is expressed in (27). This expression is further exploited to reveal the implication of the coupling is the piezoelectric material ${k_e^{piezo}}^2$ and of the stored energy ratio between the piezoelectric material and the complete structure. Considering that the energy at null electric field ($E_3 = 0$ ) is given by $U_p = \frac{1}{2} K_p r^2$, the alternative coupling can be expressed using (28). ${k_e^{piezo}}^2$ is expressed in (29).

$$k_e^2 = \frac{k^2}{1-k^2} = \frac{\Theta^2}{(K_p + K_{np}) C_p} = \frac{\Theta^2}{K_p C_p} \times \frac{K_p}{(K_p + K_{np})} \tag{27}$$

$$k_e^2 = {k_e^{piezo}}^2 \frac{U_p}{U_{np} + U_p} \tag{28}$$

$${k_e^{piezo}}^2 = \frac{\Theta^2}{K_p C_p} \tag{29}$$

As shown in equation (28), it is clear that two approaches are required to optimize the coupling coefficient of piezoelectric devices:

- Having a strong material coupling ${k_e^{piezo}}^2$. This can be achieved by using piezoelectric materials with good intrinsic properties and designing the structure to solicit them effectively,
- Maximizing the ratio of stored elastic energy in the piezoelectric material in comparison to non-piezoelectric materials $U_p/(U_{np} + U_p)$.

The objective of this work is to optimize the material coupling ${k_e^{piezo}}^2$ through the use of lateral bars. The next section outlines how lateral strain and stress considerations influence the coupling ${k_e^{piezo}}^2$ and discusses the benefits of minimizing lateral strain.

### *2.3 Plane strain and plane stress assumption*

As usually done in piezoelectric cantilever modeling and according to their dimensions [43], the stress is assumed to be zero in the z-direction ($T_3 = 0$).Furthermore, one of the two following assumptions can be considered for the lateral stress:

(i) plane stress assumption is generally used for narrow beams ($B/L_b$ < 0.2) for which no stress is considered ($T_2 = 0$) in the out-of-plane direction (Oy in Figure 1)

(ii) plane strain assumption is generally used for wide beams ($B/L_b > 5$) for which no strain is considered ($S_2 = 0$) in the out-of-plane direction.

Thanks to these assumptions and equation (12), the longitudinal stress $T_1$ and the displacement field $D_3$ in piezoelectric material on a cantilever can be modeled through in-plane constitutive equations given in (30) and (31) .

$$\begin{cases} T_1 = c_{11}^{ef} S_1 - e_{31}^{ef} E_3 & (30) \\ D_3 = e_{31}^{ef} S_1 + \epsilon_{33}^{ef} E_3 & (31) \end{cases}$$

Where $c_{11}^{ef}$, $e_{31}^{ef}$, $\epsilon_{33}^{ef}$ are the effective piezoelectric coefficients according to the given plane stress or plane strain consideration. The piezoelectric effective coefficients are deduced from the piezoelectric matrix $\boldsymbol{d}$, the compliance matrix $\boldsymbol{s^E}$ and the free dielectric matrix $\boldsymbol{\epsilon^T}$ (with $\boldsymbol{e} = \boldsymbol{d s^{E^{-1}}}$ , $\boldsymbol{c^E} = \boldsymbol{s^{E^{-1}}}$ and $\boldsymbol{\epsilon^S} = \boldsymbol{\epsilon^T} - \boldsymbol{d s^{E^{-1}} d^t}$) in equations (32) to (34).

$$c_{11}^{ef} = \begin{cases} \dfrac{1}{s_{11}^E} & \text{plane stress} \\ \dfrac{s_{22}^E}{s_{22}^E s_{11}^E - {s_{12}^E}^2} & \text{plane strain} \end{cases} \quad (32)$$

$$e_{31}^{ef} = \begin{cases} \dfrac{d_{31}}{s_{11}^E} & \text{plane stress} \\ \dfrac{s_{22}^E d_{31} - d_{32} s_{12}^E}{s_{22}^E s_{11}^E - {s_{12}^E}^2} & \text{plane strain} \end{cases} \quad (33)$$

$$\epsilon_{33}^{ef} = \begin{cases} \epsilon_{33}^T - \dfrac{d_{31}^2}{s_{11}^E} & \text{plane stress} \\ \epsilon_{33}^T + \dfrac{s_{22}^E d_{31}^2 - 2\, d_{31} d_{32} s_{12}^E + s_{11}^E d_{32}^2}{{s_{12}^E}^2 - s_{11}^E s_{22}^E} & \text{plane strain} \end{cases} \quad (34)$$

Under the assumptions on the lateral strain, the Rayleigh method presented in section 2.2. can be revised with a limited number of components. The strain and stress are 4 row vectors, given by $\boldsymbol{S} = [\, S_1 \quad S_4 \quad S_5 \quad S_6]^t$ and $\boldsymbol{T} = [\, T_1 \quad T_4 \quad T_5 \quad T_6]^t$. The piezoelectric constant vector is a 4-column vector given by $\boldsymbol{e} = \left[\, e_{31}^{ef} \quad 0 \quad 0 \quad 0\right]$, and the elastic constant matrix is a diagonal 4 × 4 matrix, given by $\boldsymbol{c^E} = \mathrm{diag}\{c_{11}^{ef}; c_{44}^E; c_{55}^E; c_{66}^E\}$. The coupling term $\Theta$, the piezoelectric stiffness $K_p$ and the capacitance $C_p$ are recalculated using equations (21) and (24) and are given in (35) to (37).

$$\Theta = \int_{V_p} N_1 e_{31}^{ef} N_v \quad (35)$$

$$K_p = \int_{V_p} N_1^2 c_{11}^{ef} + N_4^2 c_{44}^E + N_5^2 c_{55}^E + N_6^2 c_{66}^E \quad (36)$$

$$C_p = \int_{V_p} N_v^2 \epsilon_{33}^{ef} \quad (37)$$

From equations (29), (35) and (36), a simple expression of the material coupling ${k_e^{piezo}}^2$ can be obtained by considering null shear components ($N_4 = N_5 = N_6 = 0$), as usually done thin piezoelectric cantilever. This expression is given in (38), where $k_{e31}^2$ is the expedient electromechanical coupling coefficient of the piezoelectric material that depends on the effective piezoelectric coefficients $e_{31}^{ef}$, $c_{11}^{ef}$ and $\epsilon_{33}^{ef}$, as given in (39) [21]. Therefore, $k_{e31}^2$ varies according to the plane stress or plane strain assumption.

$$k_{\mathrm{e}}^{\mathrm{piezo}\ 2} = {k_{e31}}^2 \times \frac{\left(\int_{V_p} N_1 N_v\right)^2}{\int_{V_p} N_1^2 \times \int_{V_p} N_v^2} \quad (38)$$

$$k_{\mathrm{e}31}^2 = \frac{{e_{31}^{ef}}^2}{c_{11}^{ef} \epsilon_{33}^{ef}} \quad (39)$$

Equation (38) shows that ${k_e^{piezo}}^2$ in beams depends on $k_{e31}^2$ as well as on the strain and the potential distributions, through the integral of $N_1$ and $N_v$. It should be noted that if the piezoelectric material were subjected to shear, this would slightly increase $K_p$ in equation (36) and decrease ${k_e^{piezo}}^2$ in equation (29)

We observe that $k_{31}^2$, expressed in (40), can get two values depending on the aspect ratio: the coupling coefficient corresponding to the plane stress assumption, noted ${k_{31}^l}^2$, and the coupling coefficient corresponding to the plane strain assumption, noted ${k_{31}^w}^2$. They are expressed in (41) and (42) respectively from the equation (40) and equations (32) to (34).

$$k_{31}^2 = \frac{k_{e31}^2}{k_{e31}^2 + 1} = \frac{{e_{31}^{ef}}^2}{c_{11}^{ef}\epsilon_{33}^{ef} + {e_{31}^{ef}}^2} \tag{40}$$

$${k_{31}^l}^2 = \frac{d_{31}^2}{\varepsilon_{33}^T s_{11}^E} \tag{41}$$

$${k_{31}^w}^2 = \frac{\left(d_{31} - \frac{s_{12}^E}{s_{22}^E} d_{32}\right)^2}{\left(\varepsilon_{33}^T - \frac{d_{32}^2}{s_{22}^E}\right)\left(s_{11}^E - \frac{{s_{12}^E}^2}{s_{22}^E}\right)} \tag{42}$$

Most piezoelectric materials, such as ceramics as well as [001] axis-polarized PMN-PT and PZN-PT single crystals, have a higher mode 31 coupling coefficient in the plane strain configuration (${k_{31}^l}^2 < {k_{31}^w}^2$). Others, such as [011] axis-polarized PMN-PT and PZN-PT single crystals, have a higher coupling coefficient in the plane stress configuration (${k_{31}^l}^2 > {k_{31}^w}^2$). These propensities depend on crystal classes and polarization directions. Values of ${k_{31}^l}^2$ and ${k_{31}^w}^2$ for PZT and KNN ceramic materials are computed from the supplier's data sheet and provided in Table 1.

Table 1: Example of calculated material coupling coefficient from the data sheet for 2 piezoelectric ceramic materials

| | Plane stress coupling ${k_{31}^l}^2$ | Plane strain coupling ${k_{31}^w}^2$ |
|---|---|---|
| PZT-5A Noliac NCE51 [44] | 15.1 % | 34.3 % |
| KNN NiTerra LF05B [45] | 5.5 % | 10.8 % |

Wide cantilever beams therefore have to be designed to maximize the electromechanical coupling coefficient using ceramics. Piezoelectric ceramics belong to the ∞m class of crystals [46]. For this configuration, $d_{32} = d_{31}$, $s_{22} = s_{11}$, and the coupling coefficient considered in plane strain ${k_{31}^w}^2$ can be expressed by equation (43) as a function of the coupling coefficient considered in plane stress ${k_{31}^l}^2$. $\nu_p$ is a coefficient equivalent to Poisson's ratio given in (44).

$${k_{31}^w}^2 = \frac{{k_{31}^l}^2}{1 - {k_{31}^l}^2} \frac{\left(1 + \nu_p\right)^2}{1 - \nu_p^2} \tag{43}$$

$$\nu_p = -\frac{s_{12}^E}{s_{11}^E} \tag{44}$$

Since $\nu_p$ is always positive and less than 1 for ceramics, equation (43) shows that ${k_{31}^w}^2$ is always greater than ${k_{31}^l}^2$ for ceramics. Equation (43) is also true for 4 mm and 6 mm class materials when axes 1 and 3 of the piezoelectric materials correspond to the $O_x$ and $O_z$ axes of the cantilever, respectively.

This emphasizes the importance of reducing the lateral strain in piezoelectric cantilevers to maximize the electromechanical coupling coefficient. In the work [21], it has been shown that widening beams could increase the coupling by +50 %. Meanwhile, as a comparison, improving the strain distribution (e.g. by using trapezoidal cantilever) have been shown to increase only of + 33 % of the coupling in the best-case scenario compared to a rectangular cantilever with a point tip mass. In a theoretical aspect, having a Poisson ratio $\nu_p$ equal to 0.3, which is a common value for piezoelectric ceramics, the factor $\frac{(1+\nu_p)^2}{1-\nu_p^2}$ is equal to 1.85. This shows that ${k_{31}^w}^2$ is around 85 % greater than ${k_{31}^l}^2$. While achieving a configuration with zero lateral strains may be challenging, reducing it significantly impacts the coupling.

## 3. Numerical validation

### *3.1. Presentation*

This section presents a numerical study of the addition of lateral bars to a piezoelectric cantilever. Numerical simulations and experiments are performed using lead-free KNN material. While the study focuses on one material for validation, the results are relevant to all ceramic piezoelectric materials. The impact of the bars on the coupling coefficient is studied using 3-dimensional finite element modal analyses in Comsol Multiphysics. The substrate is composed of aluminium for its easiness of fabrication, and the cantilever is designed to resonate at approximately 30 Hz for general vibration energy harvesting applications. The materials and dimensions of the initial cantilever are defined in Table 2. The selected KNN material LF05B is a hard-ceramic material with a high quality factor and weak nonlinearity, making it well suited for vibration energy harvesting applications.

Table 2 : Properties of the studied cantilever

| | |
|---|---|
| Lead free piezoelectric material | LF05B from Niterra (KNN based) |
| Substrate material | Aluminum |
| Proof mass material | Steel |
| Piezoelectric thickness $h_p$ | 0.4 mm |
| Substrate thickness $h_s$ | 1.5 mm |
| Piezoelectric and beam length $L_b$ | 70 mm |
| Beam and mass width $B$ | 15 mm |
| Mass length $L_m$ | 60 mm |
| Mass height $H_m$ | 15.5 mm |

The material coefficients of the piezoelectric material are computed from an impedance analysis measurement of the piezoelectric plates and using the method given in literature [47,48]. These parameters are given in Table 3.

Table 3 : Measured properties of the KNN based LF05B piezoelectric material

| | $d_{31}$ (pm.V$^{-1}$) | $s_{11}^E$ (×10$^{-12}$Pa$^{-1}$) | $\epsilon_{33}^T$ (F.m$^{-1}$) | $\rho_p$ (kg.m$^{-3}$) |
|---|---|---|---|---|
| LF05B | -42.7 | 9.1 | 574 $\epsilon_0$ | 4 500 |

A modal analysis is performed on COMSOL Multiphysics to determine the short circuit and open circuit resonance frequencies of the first mode of the “initial” cantilever, without bars. A second simulation is performed in which 8 steel bars of dimensions $l_b \times h_b \times B = 2 \times 4 \times 15\ mm^3$ (4 on each side of the beam, arranged in a staggered pattern) are added as shown in Figure 4. For this simulation, the bars are considered perfectly fixed to the piezoelectric material. The distributions of the longitudinal and lateral deformations of the mode at short circuit condition are given in Figure 5 and Figure 6.

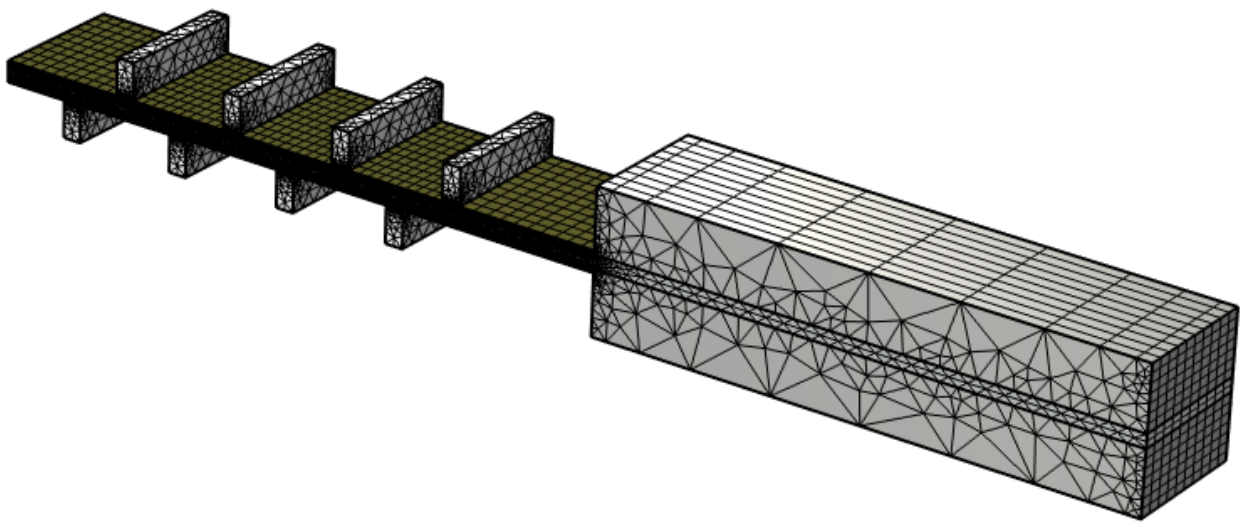

Figure 4: Meshed cantilever on COMSOL Multiphysics

The longitudinal and lateral strain distributions are given in Figure 6. We notice that the longitudinal strain is preserved between the “initial” cantilever and the “enhanced” cantilever with bars while the lateral strains are reduced. The short circuit resonance frequency ($f_{sc}$) and the open circuit resonance frequency ($f_{oc}$) are used to calculate the coupling coefficient using equation (45). The resonant frequency and the coupling coefficient of the “initial” cantilever are also computed with 2D simulations to determine the boundary values of plane stress and plane strain configurations. The values are reported in Table 4.

$$k^2 = \frac{f_{oc}^2 - f_{sc}^2}{f_{oc}^2} \qquad (45)$$

The short-circuit resonance frequency is slightly influenced by the addition of bars while the coupling coefficient is increased by more than 30 % compared to the initial configuration. This demonstrates the potential of lateral bars with a ceramic piezoelectric material. The coupling coefficients computed in 2D with plane stress and plane strain assumptions provide useful indications of this increase. While the coupling of the 3D initial cantilever is close to the one computed with 2D simulations under plane stress assumption in Table 4, the device with bars ($k^2 = 3.7$ %) goes closer to the coupling computed with the plane strain configuration. As lateral strain is not fully nullified, and the energy distribution in the cantilever is affected, we are unable to reach the full value of the coupling of the plane strain configuration.

The coupling coefficient of the piezoelectric material when integrated into the cantilever, $k_{piezo}^2$, along with the ratio of the elastic energy stored in the piezoelectric material to that stored in the entire system, provides insight into the influence of the bars on the cantilever's behavior. Using the method described in [30], the material coupling $k_{piezo}^2$ can be determined from equation (46) , where $U_{piezo}$ and $U_{tot}$ are the global elastic energy and the elastic energy of the piezoelectric elements, respectively, computed using short-circuit modal analysis. The material coupling $k_{piezo}^2$ and the energy ratio $U_{piezo}/U_{tot}$ are also given in Table 4 for 2D and 3D analyses. It is observed that the addition of bars does not influence much the energy ratio as it goes from 78 % to 75 %. Nevertheless, the material coupling increases significantly: The coupling $k_{piezo}^2$ of the device with bars is equal to 5.4 % while it is equal to 3.8 % without bars. It does not reach the plane strain coupling of 6.8 % due to residual lateral strain and/or strain inhomogeneity in the material. As a matter of illustration, the $k_{piezo}^2$ with the 2D plane strain configuration is not equal to ${k_{31}^w}^2$ from Table 1 as the strain is not uniform in the piezoelectric elements.

$$\frac{k_{piezo}^2}{1 - k_{piezo}^2} = \frac{k^2}{1 - k^2} \frac{U_{tot}}{U_{piezo}} \qquad (46)$$

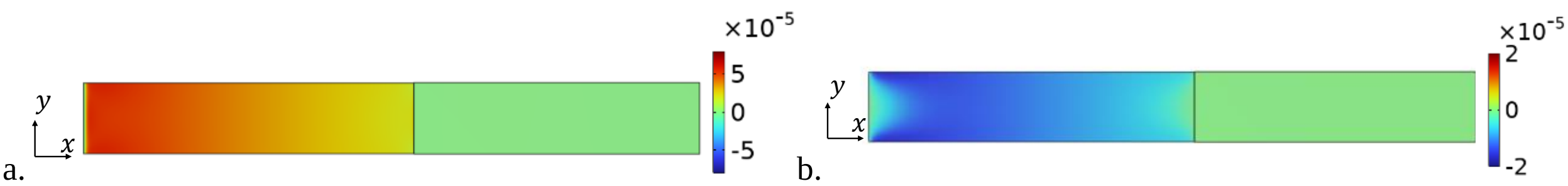

Figure 5: a. Longitudinal strain $S_1$ of the "initial" cantilever seen from above for a tip displacement of 0.1 mm at $x = L_b$, b. Lateral strain $S_2$

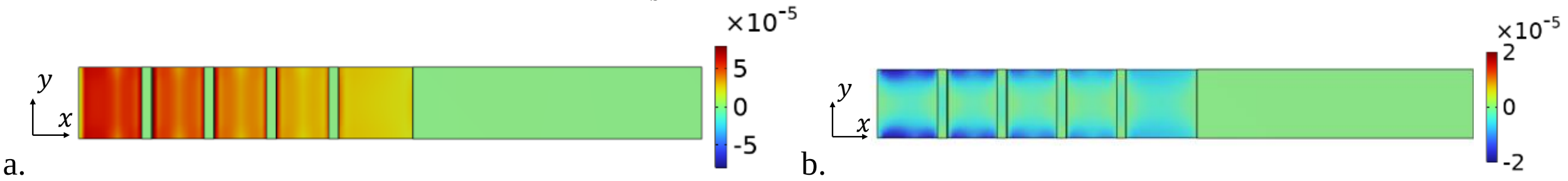

Figure 6: a. Longitudinal strain $S_1$ of the "enhanced" cantilever seen from above for a tip displacement of 0.1 mm at $x = L_b$, b. Lateral strain $S_2$. The four bars on each side of the beam are staggered.

Table 4 : Influence of adding bars determined by simulations

| Cantilever type | "Initial" 2D plane stress | "Initial" 2D plane strain | "Initial" 3D | "Enhanced" with bar 3D |
|---|---|---|---|---|
| Short-circuit resonant frequency $f_{sc}$ | 32.1 Hz | 33.6 Hz | 32.4 Hz | 34.0 Hz |
| Global coupling $k^2$ | 2.67 % | 5.01 % | 2.89 % | 3.77 % |
| $U_{piezo}/U_{tot}$ | 78 % | 77 % | 78 % | 75 % |
| Material coupling $k_{piezo}^2$ | 3.5 % | 6.8 % | 3.8 % | 5.4 % |

To have a better visualization of the strain, Figure 7 depicts the longitudinal and lateral strains $S_1$ and $S_2$ at the center of the upper piezoelectric patch for a displacement of 0.1 mm at the tip end of the beam (at $x = L_b$). While the lateral strain is significantly reduced, the longitudinal strain is not much affected. As having a homogeneous longitudinal distribution of $S_1$ improves $k^2$ in cantilevers [21], maintaining a good longitudinal distribution despite the bars is necessary to maximize the coupling. As the proposed solution has only a minor impact on the longitudinal strain distribution, the use of lateral bars could be combined with strategies to homogenise the strain, such as the use of triangular beams. As a long proof mass is used here to homogenise the strain, the use of a triangular beam would not provide a significant further improvement in the coupling [21].

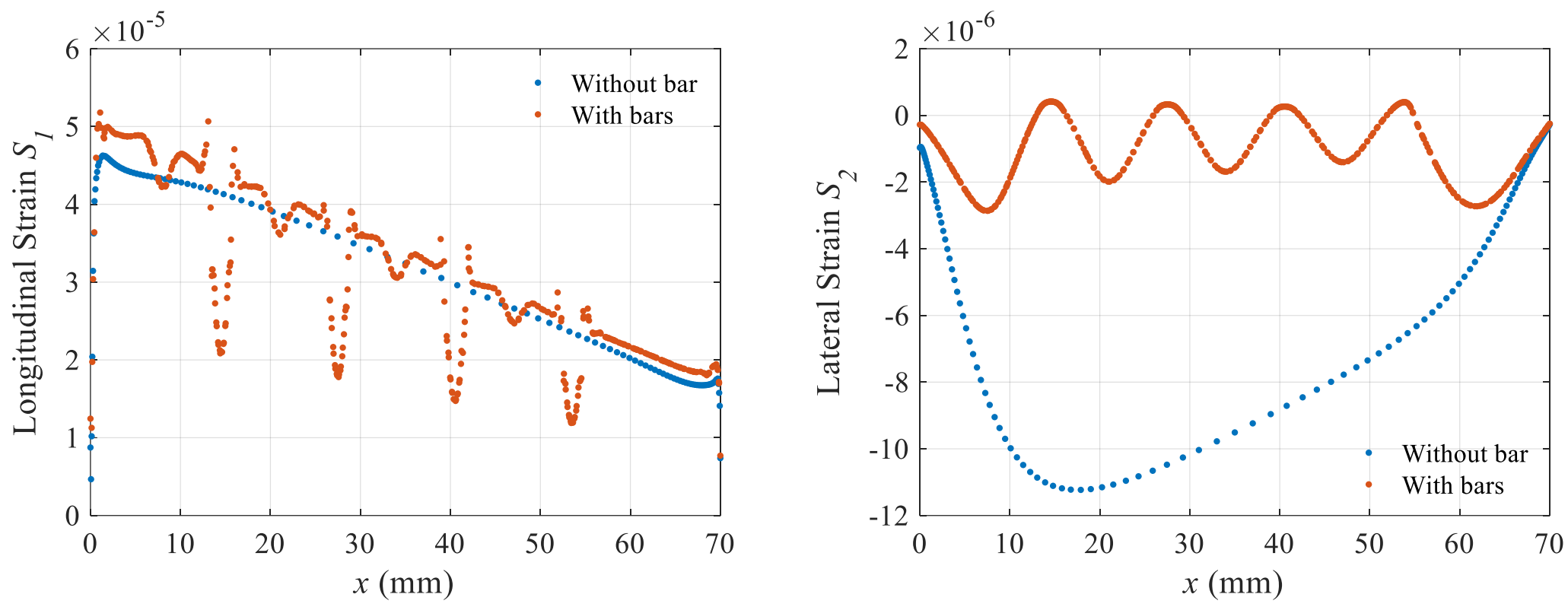

Figure 7: Longitudinal strain $S_1$ on the left and lateral strain $S_2$ on the right simulated by COMSOL 3D at the center of the upper piezoelectric patch for a tip displacement of 0.1 mm at $x = L_b$

### *3.3 Parametric study*

While the optimal dimensions of the bars depend on a number of factors, including the materials and the number of bars, it is not possible to analyze all of these factors. Instead, the impact of bar dimensions for a given configuration is here examined, with a view to providing some design recommendations. Further parametric studies are provided in Appendix A. The parametric study has been realized thanks to COMSOL Multiphysics. The original beam, whose parameters are given in Table 2 is unchanged. The bars remain in steel and their location are unchanged. Their length $l_b$ and height $h_b$ are varied. Figure 8 represents the computed squared coupling coefficient $k^2$ using equation (45) as a function of $l_b$ for 5 values of $h_b$ and for the "initial" cantilever Figure 9 represents the energy ratio $U_{piezo}/U_{tot}$ and material coupling $k^2_{piezo}$ using (46).

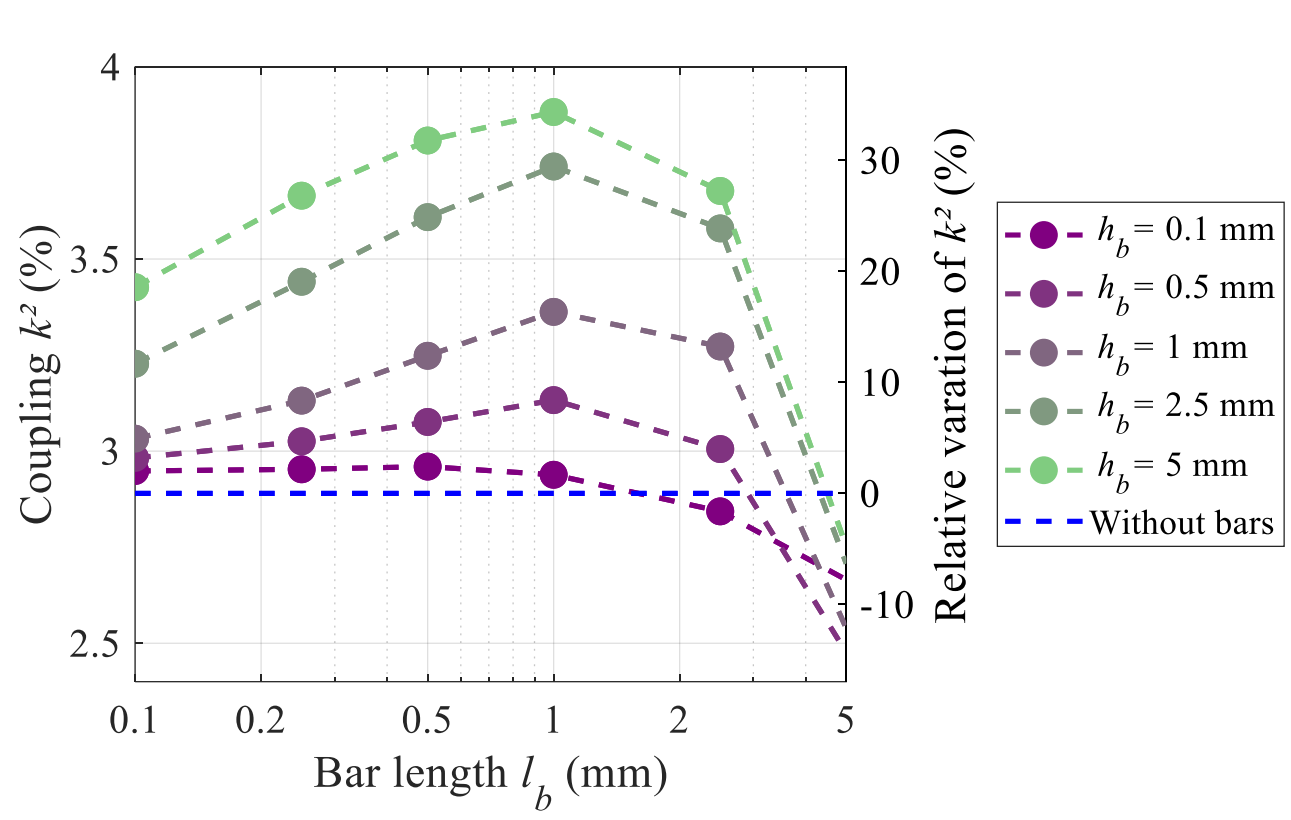

Figure 8: Simulated squared coupling coefficient $k^2$ as a function of bars dimensions. The relative variation compared to the one without bars is given in the right axis

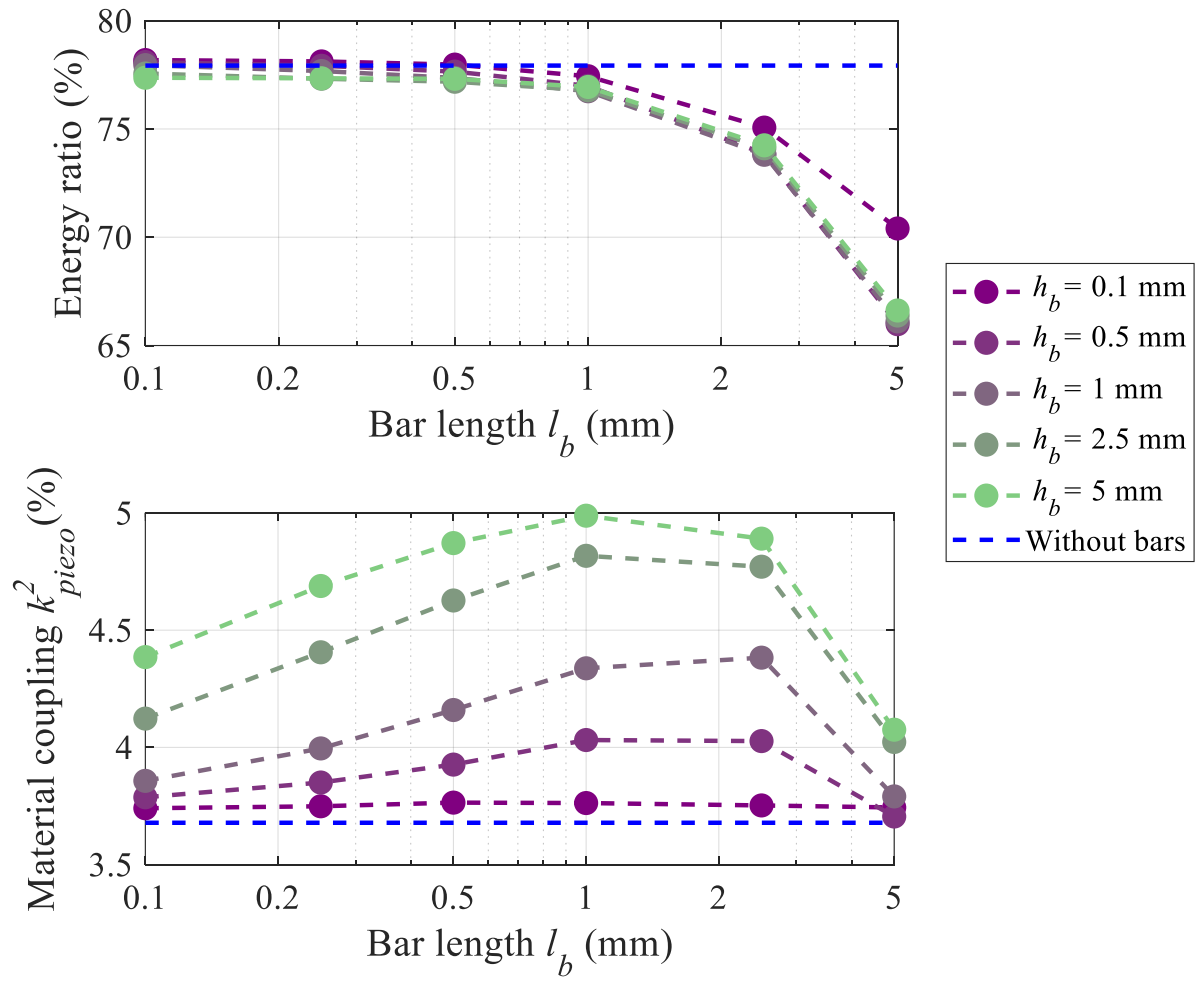

Figure 9: Energy ratio $U_{piezo}/U_{tot}$ and material coupling $k^2_{piezo}$ as a function of bars dimensions, compared to the one in the "initial" cantilever

From Figure 8 and Figure 9, it is notices that higher $h_b$ gives more influence of the bars on the coupling. Globally, the coupling $k^2$ increases with $h_b$. Having stiffer bars therefore allows to better prevent the lateral strain in the beam and get a better material coupling, which tends to ${k_{31}^{w}}^2$. In the same manner, the coupling is firstly improved with the increase of the bars' length $l_b$ as the influence of the bar increases. However, a decrease is observed in the coupling for $l_b$ superior to 1 mm. This decrease is attributed to the blocking of the longitudinal strain $S_1$ and the decrease of the energy ratio $U_{piezo}/U_{tot}$. Indeed, with higher $l_b$, the influence of the beam stiffness on the beam becomes too significant, preventing it from bending. The inhomogeneity of the longitudinal strain, such as that shown in Figure 7 also becomes too significant, and the coupling $k^2_{piezo}$ decreases. In Figure 8, it can be even noticed that the coupling with the bars becomes less than that of the initial beam without the bars. This therefore shows that there is an optimal length $l_b$ to maximize the coupling coefficient $k^2$.

## 4. Experimental validation

This section presents the experimental validation of the use of lateral bars. An initial aluminum cantilever with KNN piezoelectric ceramic is designed, fabricated and tested under vibration. Then steel bars are added thanks to epoxy glue and the final prototype is tested to analyze the influence of the bars results. The resulting increase in $k^2$allows for a wider tuning range of the resonant frequencies using various resistive loads.

### *4.1. Fabrication and identification*

The LF05B KNN based piezoelectric ceramic plates have been bonded with epoxy glue (3M DP460) in our laboratory to the aluminum beam. The geometrical parameters are the one of Table 2. The steel proof mass has been also bonded to the substrates with the epoxy glue.  Figure 10 depicts the final prototype. The power output of the prototype has been determined under sinusoidal vibration excitation at 0.1 m/s² amplitude using the test bench represented in Figure 11. The vibrations are generated and controlled by an electromagnetic vibrator (K2075E-HT) driven by a dSpace board and the acceleration level is monitored using an accelerometer (Piezotronics 333B40). The dSpace board is controlled by a dedicated Matlab script that defined the frequency, the acceleration level and controlled the programmable electrical resistance. Experiments have been performed for 12 resistive loads logarithmically spaced between 20 kΩ and 2 MΩ over 60 excitation frequencies linearly spaced between 25 Hz and 26.8 Hz. The displacement is  measured on the proof mass close to the beam end thanks to a laser vibrometer as shown in Figure 11.

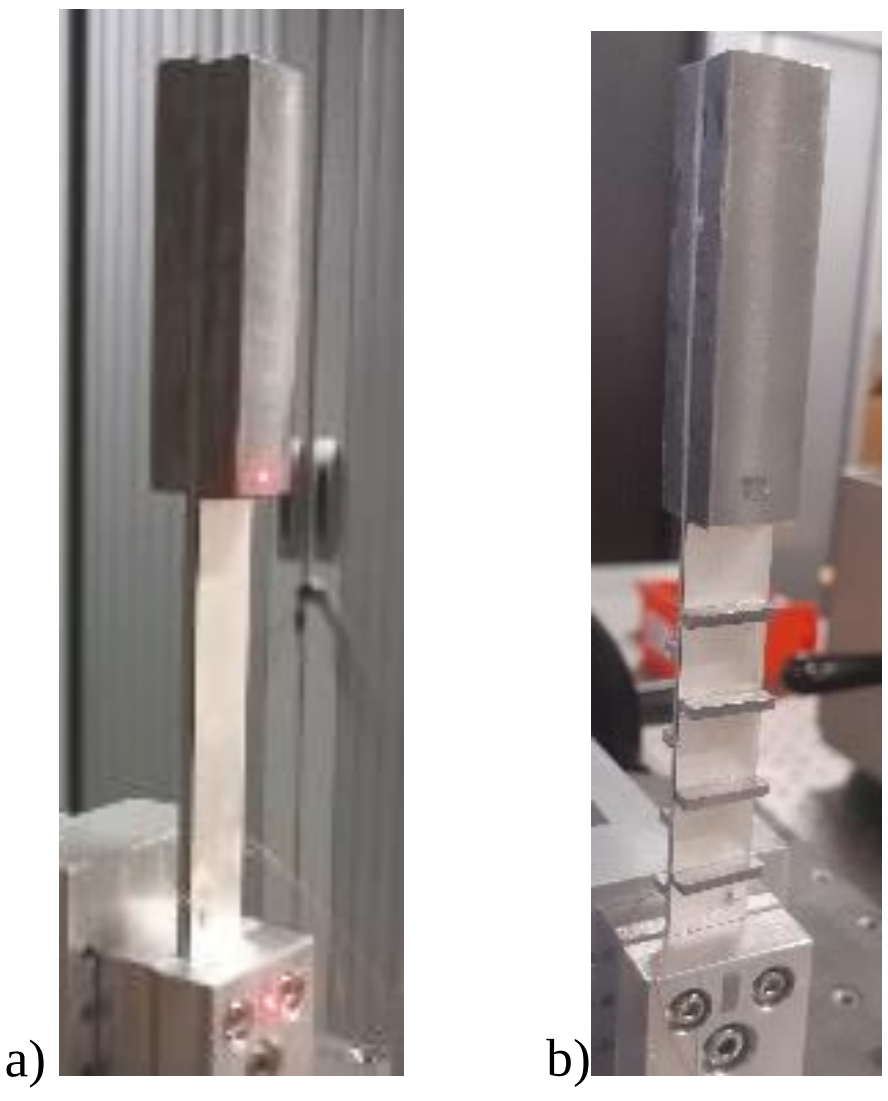

Figure 10: Picture of the prototype on the shaker a) without bars and b) with bars

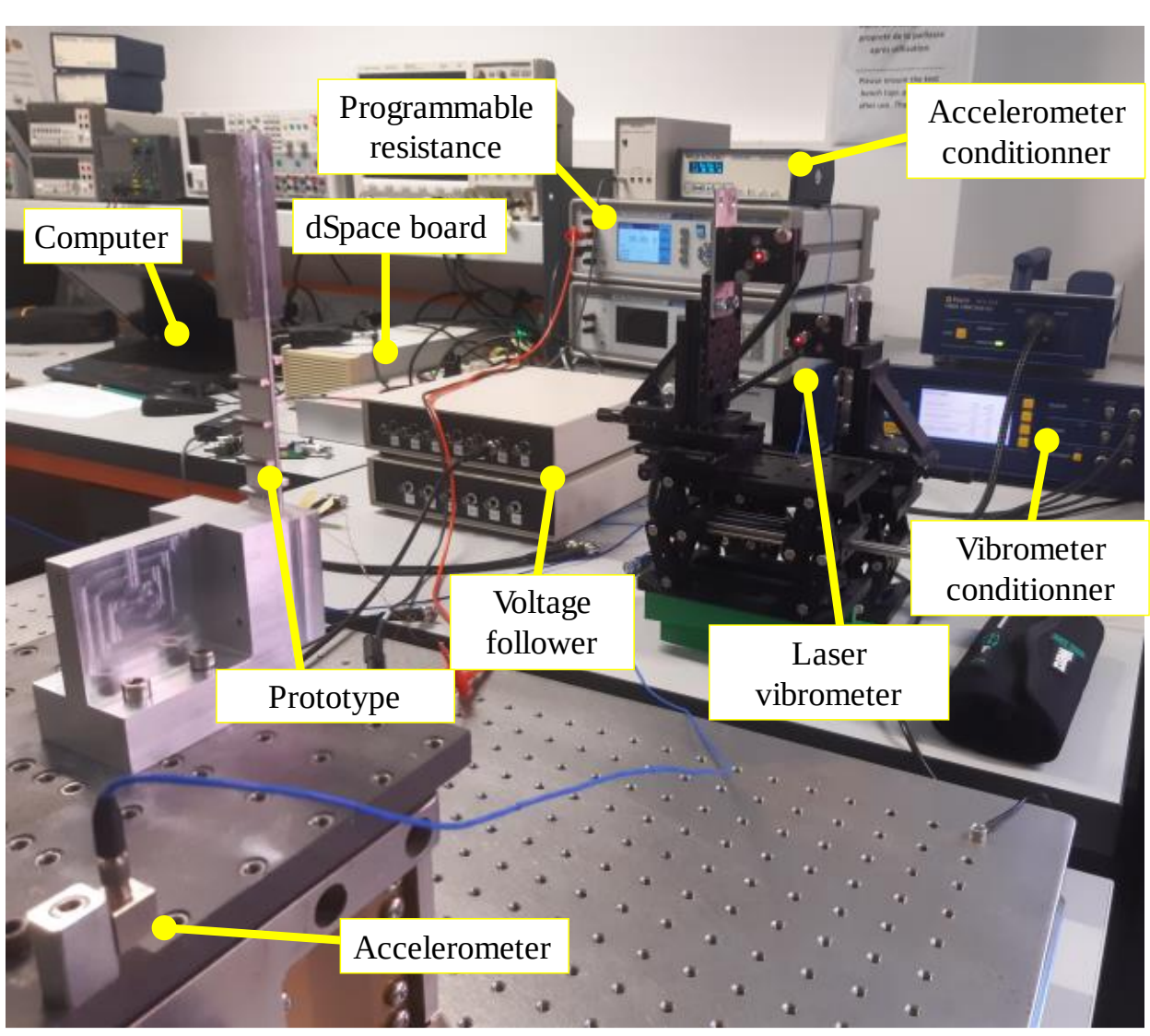

Figure 11: Experimental test bench

The displacement amplitude and mean power have been fitted with the one mechanical degree of freedom linear model presented in [49], from which the equations of motion and current are given in equation (47). $r$, $v$ and $w_b$ are the relative displacement of the proof mass, the voltage across the electrodes and the displacement of the base, respectively. The coefficients definition and the determined values are summarized in Table 5. The mean harvested power and the displacement amplitude as well as the fitted model are shown in Figure 12.

$$\begin{cases} \ddot{r} + \frac{\omega_0}{Q_m}\dot{r} + \omega_0^2 r - \frac{\Theta}{\mathrm{M}} v = -\frac{B_f}{M}\ddot{w}_B \\ \Theta\dot{r} + C_p\dot{v} + \frac{v}{R_{load}} = 0 \end{cases} \tag{47}$$

Table 5 : Coefficients deduced from the experiments.

| Name | Symbol | Prototype without bars | Prototype with bars |
|---|---|---|---|
| Resonant pulsation | $\omega_0$ | 162 rad/s | 167 rad/s |
| Equivalent mass | $M$ | 222 g | 230 g |
| Forcing term | $B_f$ | 142 g | 151 g |
| Coupling term | $\Theta$ | -1.8×10$^{-3}$ N/V | -2.2×10$^{-3}$ N/V |
| Capacitance | $C_p$ | 30.9 nF | 30.6 nF |
| Quality factor | $Q_m$ | 81 | 66 |

Four 2 × 4 × 15 mm$^3$ steel bars were bonded to each side of the beam in a staggered pattern using 3M DP460 epoxy adhesive. The "enhanced" cantilever prototype with bars is shown in Figure 10.b. It has also been tested under a 0.1 m/s² amplitude sinusoidal vibration. The cantilever is tested with the same resistive loads as the "initial" cantilever and the excitation frequencies are linearly spaced between 26 Hz and 28 Hz in order to stay around the resonance frequency. The equivalent parameters of the "enhanced" cantilever are determined with a fit using equation (47) and are given in Table 5. The harvested mean power and displacement amplitude measurements are shown in Figure 12.

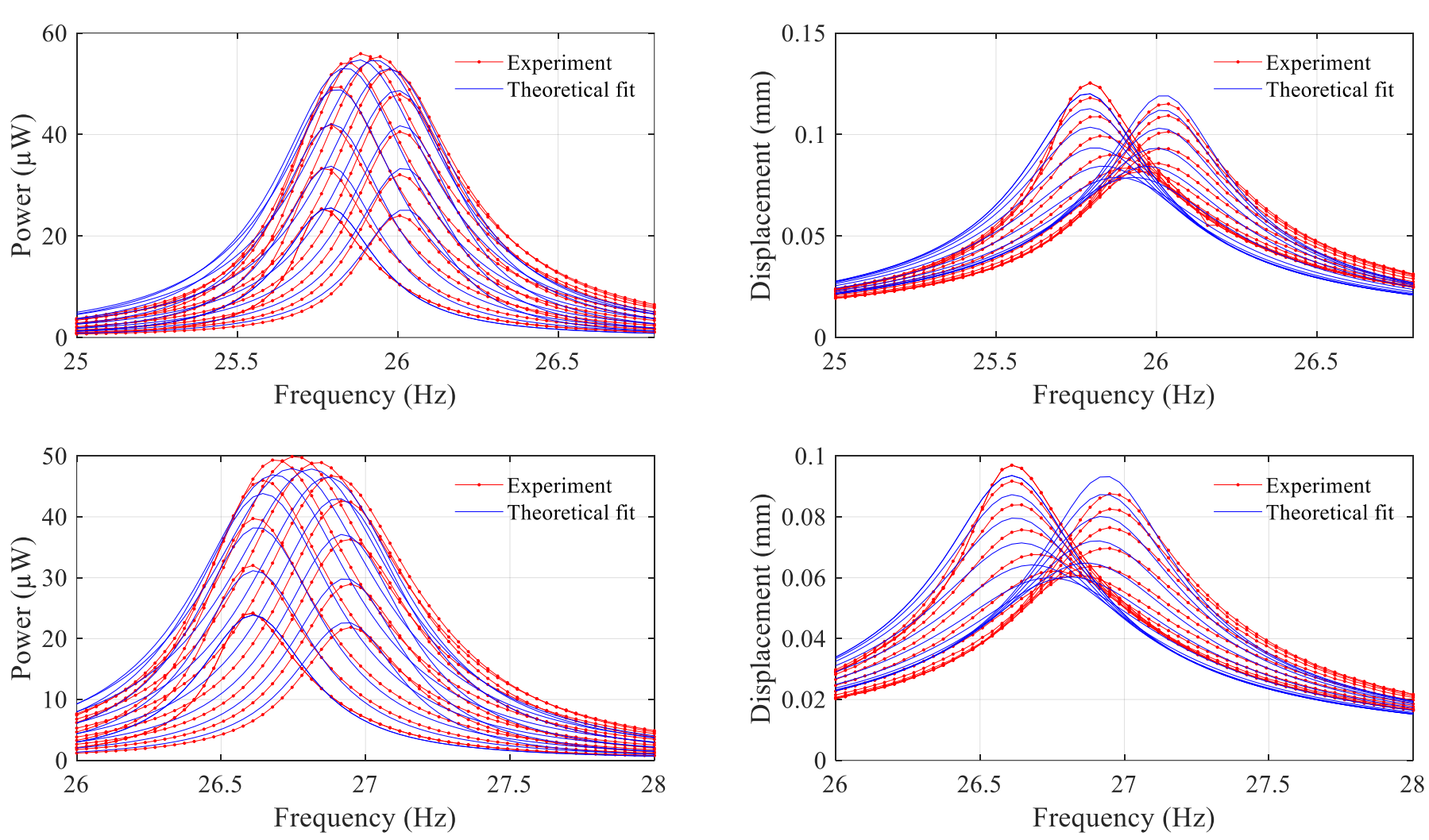

Figure 12: Measured mean power and displacement amplitude of the "initial" prototype without bars (top) and the "enhanced" prototype with glued bars (bottom). The acceleration amplitude is $|\ddot{w}_b| =$ 0.1 m/s²

From Table 5, the short-circuit resonant frequency $f_{sc}$ and electromechanical coupling coefficient $k^2$ have been calculated using equation (48), where $\theta$, $K$ and $C_p$ are coupling term, the stiffness and the capacitance. The stiffness is obtained using $K = \omega_0^2 M$. The expected power using a tunable resistive load for resonant frequency tuning is determined from the envelop of the measured power for both prototypes and is represented in Figure 13. The expected frequency bandwidth at half of the maximal harvested power, $\Delta f$, is also represented and given. The resonance frequency, the coupling $k^2$, the maximal power and the relative bandwidth to the resonance frequency $\Delta f / f_{sc}$ are reported in Table 6.

$$k^2 = \frac{\Theta^2}{K C_p + \Theta^2} \qquad (48)$$

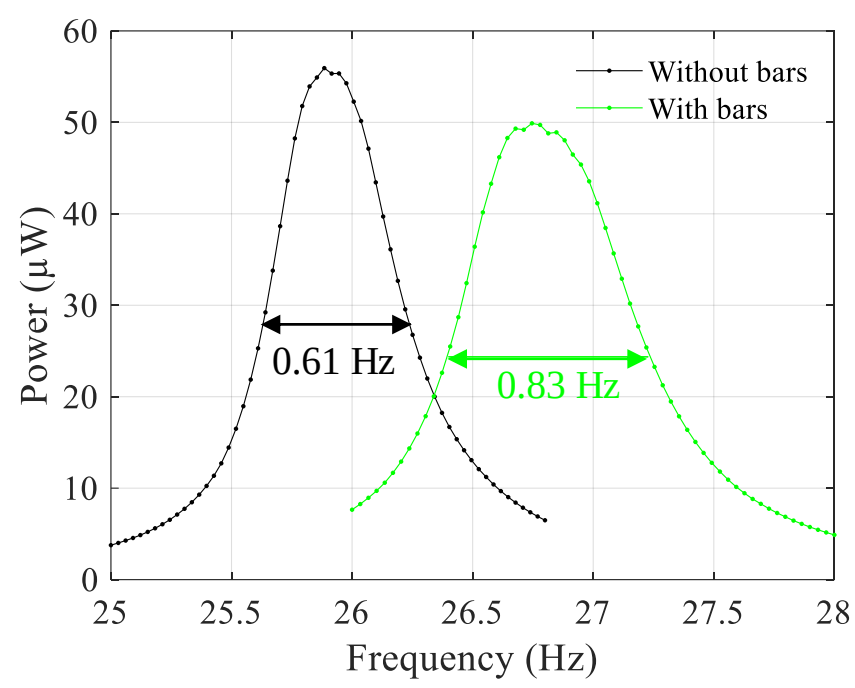

Figure 13: Expected power using a tunable resistive load for resonant frequency tuning deduced from the measured power

We notice that the coupling coefficient $k^2$ passes from 1.9 % to 2.5 % by glueing the bars on the prototypes. This represents a relative increased of more than 30 %. As a comparison, the resonant frequency, the equivalent mass $M$, forcing term $B_f$ and capacitance $C_p$ are almost unchanged, according to Table 5. The FEM simulations predict the increase of the coupling through the bar additions correctly, as the expected and experimental increases of $k^2$ are both almost equal to 30 % in Table 4 and Table 5. The resonant frequencies and coupling coefficient values obtained by FEM differ slightly from measurements and are explained by the glue between the substrate and the piezoelectric material as well as the spacing between the piezoelectric ceramics with the clamp end and the proof mass. The latter are not considered in FEM while they can have a significant impact on the coupling. While the values cannot be accurately determined, previous works reported coupling that can be divided by 2 due to spacing of few hundreds of micrometers and glue thicknesses of few dosen of millimeters [9,21].

Table 6 : Performances comparisons

| Parameter | Prototype without bar | Prototype with bars |
|---|---|---|
| Frequency $f_{sc}$ | 25.8 Hz | 26.6 Hz |
| Coupling $k^2$ | 1.9 % | 2.5 % |
| Maximal power | 55.9 µW | 49.9 µW |
| Bandwidth | 2.35 % | 3.10 % |

A decrease of the maximal power of 11 % is noticed in Table 6 with the addition of the bars, from 55.9 µW to 49.9 µW. It is imputed to the use of glue, which is a lossy material, that induces the decrease of equivalent quality factor observed in Table 5. As the quality factor of the piezoelectric material ($Q_m = 500$ according to the data sheet [45]) and the one of the initial cantilever are high, the addition of glue has a non-negligible impact. Such a behavior would be less perceptible with soft piezoelectric materials which present higher losses than hard ones or with another way to fix the bars on the piezoelectric elements. Nevertheless, the high quality factor of the devices allows for having high normalized power density (NPD) compared to the literature [9,12]. Using equation (49), the NPD of the harvesters are equal to 18.5 mW/cm$^3$/G$^2$ and 16.5 mW/cm$^3$/G$^2$ for the prototype without and with bars, respectively.

$$\mathrm{NPD} = \frac{P_{max}}{|\ddot{w}_b|^2[(L_b + L_m)H_m B]} \tag{49}$$

A wider frequency bandwidth is observed with the “enhanced” cantilever compared to the “initial” one in Figure 13. The frequency bandwidth goes from 0.61 Hz to 0.81 Hz, which corresponds to a relative frequency of 2.35 % and 3.10 % respectively. This is made possible by the increase in the coupling coefficient resulting from the addition of the bars blocking the lateral strain. Therefore, although the maximum power decreases by 11%, the bandwidth increases by 32%. The product of maximum power and bandwidth, which is used to compare vibration energy harvesters [9], increases from 131 µW.% to 155 µW.%. This shows a significant boost in the harvester's performance by the addition of bars.

In conclusion, our solution allows a significative improvement of the electromechanical performances of piezoelectric harvesters by benefiting of a better material coupling. A high-performance lead-free harvester could be designed with an electromechanical coupling coefficient $k^2$ 2.5 % and quality factor of 66 (see Table 5 and Table 6).

## Conclusion

This work presented an innovative design that improves the global electromechanical coupling coefficient $k^2$ of piezoelectric cantilevers. The electromechanical coupling of these cantilevers is enhanced by minimizing the lateral strain. This innovative design involves placing several rigid bars across the piezoelectric elements in the widthwise direction, with the aim of blocking lateral strain while allowing longitudinal strain. Using COMSOL Multiphysics simulations, it has been shown that placing convenient bars allows maximizing the coupling coefficient of the material $k^2_{piezo}$, and therefore the global coupling coefficient $k^2$.

A prototype was designed with a hard ceramic lead free KNN glued on an aluminum beam to experimentally validate the solution. The harvester was tested under vibration before and after fixing eight steel bars on the piezoelectric material. The experiment results show a significant relative increase of $k^2$, by 30% are added to the cantilever, from 1.9% to 2.5%. These improvements allow for enhanced tuning capabilities of the resonant frequency through electrical methods. Experiments involving variable resistive loads have indicated an increase in relative frequency bandwidth of up to 32%. The harvester exhibits a high normalized power density of 16.5 mW/cm³/G² by harvesting 49.9 µW at an acceleration amplitude of 0.1 m/s². Furthermore, with a relative frequency bandwidth of 3.1%, it is a highly efficient energy harvester made from lead-free material. Even though the method has been validated for KNN materials, it is fully transferable to other materials. Such results pave the way for the design of high-performance piezoelectric devices with strong coupling coefficients.

In this work, the bars were equally spaced and identical in terms of their material and geometry. Future work could involve studying the independent optimisation of the bars and their location. Furthermore, work could be undertaken using topological optimisation.

## Acknowledgements

The authors would like to acknowledge the technical support of Blaise Girard for the manufacturing of the prototypes and Louison Tourtelier-Gallo for the preliminary study. This work was supported by the French National Agency for research (ANR) through the project ANR-22-CE51-0030 (ANR Player One).

## Appendix A

This appendix provides supplementary parametric analyses that complement those carried out in section 3.3. Three studies have been carried out: the effect of beam width, the use of aluminium for the bars, and the use of four bars. The initial parameters are given in Table 2 and the bars configuration is the one presented in section 3.1. Further parametric studies involving a PMN-PT-based cantilever can be found in [33].

Figure 14 shows the coupling coefficient of the cantilever as a function of bar height for five values of beam width B, and demonstrates that increasing the beam width increases the coupling without bars. Therefore, the initial state, before the bars are added, is closer to the plane strain state. Consequently, the relative increase in the coupling coefficient is limited when the beam is wide. For $B = 15\ mm$, adding the bars increases the coupling coefficient from 2.89% to 3.90%, while for $B = 60\ mm$, it increases from 3.87% to 4.67%. Nevertheless, Figure 14 shows that a large width combined with the addition of bars provides the best coupling.

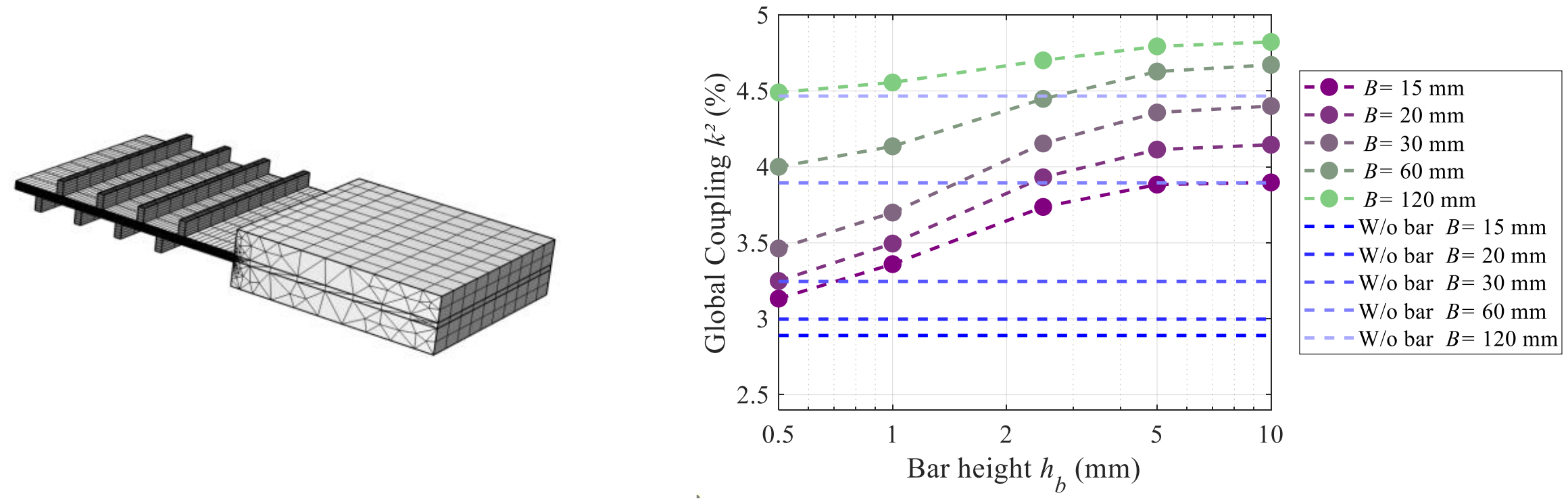

Figure 14: a. Cantilever with bars for $B = 60\ mm$, b. Coupling coefficient $k^2$ as function of bar height for several values of width for $l_b = 2\ mm$

Figure 15 shows the coupling coefficient of the cantilever as a function of the bar length $l_b$ for 3 values of bar heights $h_b$, when aluminium bars are used. The figure provides an analysis of the impact of the stiffness of the material used for the bars, given that the Young modulus of aluminium is almost three times lower than that of steel ($E_{steel} = 200\ GPa$ and $E_{Al} = 69\ GPa$). Figure 15 shows thatt aluminium exhibits similar tendencies to those observed for steel in Figure 8. However, the increase in coupling is lower when aluminium is used for low values of $l_b$. For example $k^2$ is equal to 3.51 % and to 3.74 % for $l_b = 1\ mm$ and $h_b = 2.5\ mm$, for aluminium and steel respectively. Furthermore, the optimal length $l_b$ differs between steel and aluminium. A higher value of length $l_b$ is necessary for aluminium to block the lateral strain.

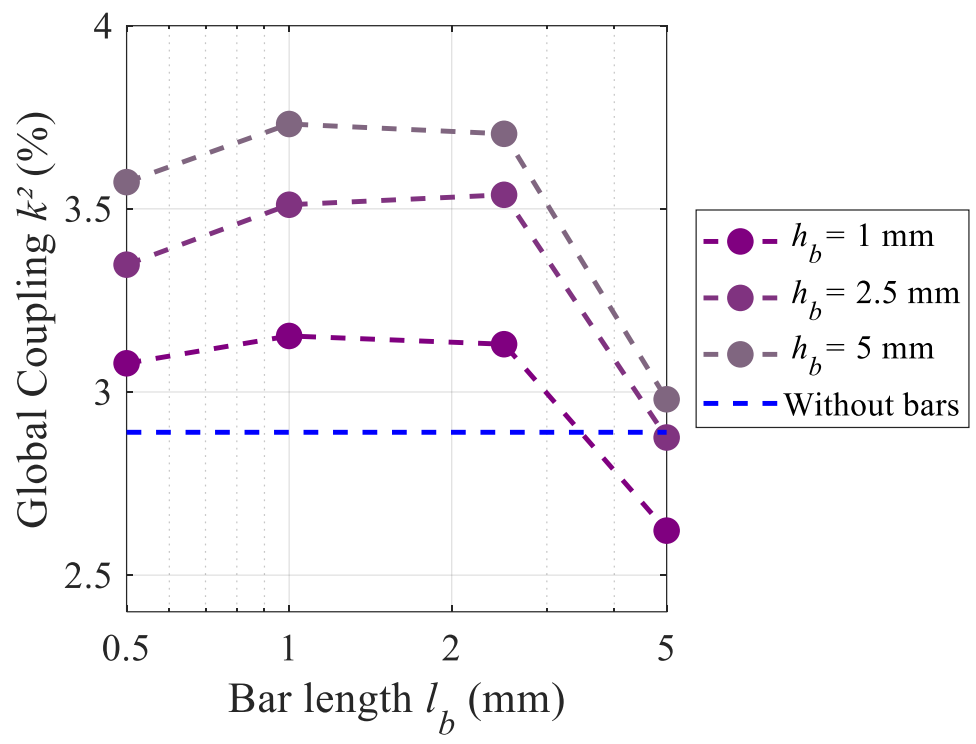

Figure 15: Coupling coefficient $k^2$ for 4 aluminum bars on each side of the beam

Figure 16 shows the coupling coefficient of the cantilever with 2 steel bars on each side of the beam as a function of the bar length $l_b$ for 4 values of bar height $h_b$. Although the coupling coefficient is lower than that achieved with four bars on each side of the beam in Figure 8, Figure 16 shows that using only 4 bars in total improves the coupling from 2.89 % to 3.60 %. As with 8 aluminium bars, the optimal bar length differs with 4 steel bars than with 8 steel bars.

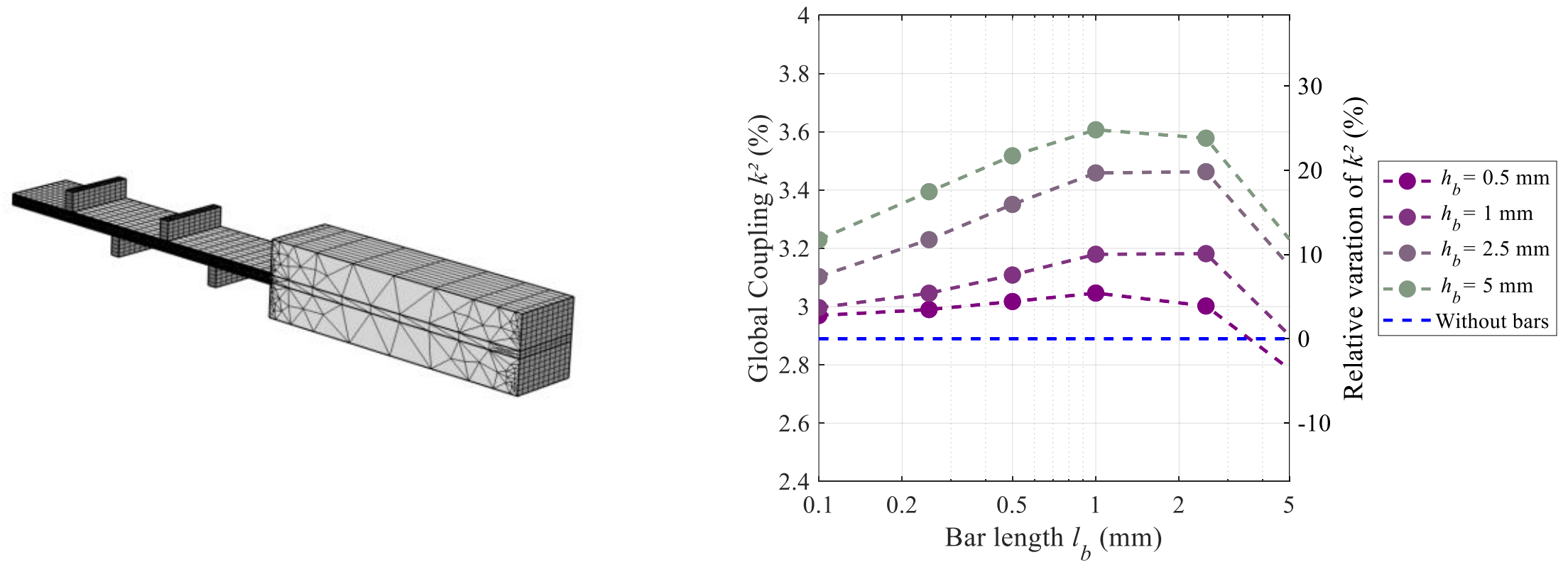

Figure 16: a. Cantilever with 2 steel bars on each side of the beam,
b. Coupling coefficient $k^2$ for 2 steel bars on each side of the beam

To conclude, these parametric studies show the following major results: i. Adding bars becomes less interesting if the width-to-length ratio B/L of the cantilever is high, ii. The Young modulus and the number of bars affect the optimal size of the bars.